%% file: main.tex
\documentclass[letterpaper]{article} % DO NOT CHANGE THIS
\usepackage[]{format/aaai23} 

\usepackage{hyperref}
\usepackage{times}  % DO NOT CHANGE THIS
\usepackage{helvet}  % DO NOT CHANGE THIS
\usepackage{courier}  % DO NOT CHANGE THIS
\usepackage{graphicx} % DO NOT CHANGE THIS
\usepackage{natbib}  % DO NOT CHANGE THIS AND DO NOT ADD ANY OPTIONS TO IT
\usepackage{caption} % DO NOT CHANGE THIS AND DO NOT ADD ANY OPTIONS TO IT
\usepackage{algorithm}
\usepackage{algorithmic}

\usepackage{xcolor}

\begin{document}

%%
%% The "title" command has an optional parameter,
%% allowing the author to define a "short title" to be used in page headers.
\title{On Safe and Usable Chatbots for Promoting Voter Participation}
% \title{BlueSky Paper: Towards Causally Grounded Rating of Multimodal AI Systems to Communicate Trust}

% \author {
%     % Authors
%     Anonymous
%     % Later - Bharath, Vishal, Kausik, Shuge, Biplav, Brett, Andrea and Vignesh
%     %All Contributors\textsuperscript{\rm 1}
% }
% Bharath Muppasani, Vishal Pallagani, Kausik Lakkaraju, Shuge Lei, Biplav Srivastava, Brett Robertson, Andrea Hickerson and Vignesh Narayanan
\author {
    % Authors
    Bharath Muppasani
    \textsuperscript{\rm 1},
    Vishal Pallagani
    \textsuperscript{\rm 1},
    Kausik Lakkaraju
    \textsuperscript{\rm 1},
    Shuge Lei
    \textsuperscript{\rm 1}, \\
    Biplav Srivastava
    \textsuperscript{\rm 1},
    Brett Robertson
    \textsuperscript{\rm 2},
    Andrea Hickerson 
    \textsuperscript{\rm 3}, 
    Vignesh Narayanan
    \textsuperscript{\rm 1}
}
\affiliations {
    % Affiliations
    \textsuperscript{\rm 1} AI Institute, University of South Carolina, USA\\
    \textsuperscript{\rm 2} School of Journalism and Mass Comm., University of South Carolina, USA \\
    \textsuperscript{\rm 3} University of Mississippi, USA \\

% biplav.s@sc.edu
}
% \affiliations {
%     % Affiliations
%     Anonymous Organization During Review
% }

%%
%% This command processes the author and affiliation and title
%% information and builds the first part of the formatted document.

\maketitle

\begin{abstract}

Chatbots, or bots for short, are multi-modal collaborative assistants that can help people complete useful tasks. Usually, when chatbots are referenced in connection with elections, they often draw negative reactions due to the fear of mis-information and hacking. Instead, in this work, we explore how chatbots may be used to promote voter participation in vulnerable segments of society like senior citizens and first-time voters. In particular, we have built a system that amplifies official information while personalizing it to users' unique needs transparently (e.g., language, cognitive abilities, linguistic abilities). The uniqueness of this work are (a) a safe design where only responses that are grounded and traceable to an allowed source (e.g., official question/answer) will be answered via system's self-awareness (metacognition), (b) a do-not-respond strategy that can handle customizable responses/ deflection, and (c) a low-programming design-pattern based on the open-source Rasa platform to generate chatbots quickly for any region. Our current prototypes use frequently asked questions (FAQ) election information for two US states that are low on an ease-of-voting scale, and have performed  initial evaluations using focus groups with senior citizens. Our approach can be a win-win for voters, election agencies trying to fulfill their mandate and democracy at large.

% Chatbots, or bots for short, are multi-modal collaborative assistants that can help people complete useful tasks. Usually, when chatbots are referenced in connection with elections, they often draw negative reactions due to the fear of misinformation and hacking. Instead, in this paper, we explore how chatbots may be used to promote voter participation in vulnerable segments of society like senior citizens and first-time voters. In particular, we build a system that amplifies official  information while personalizing it to users’ unique needs transparently. We discuss its design, build  prototypes with frequently asked questions (FAQ) election information for two US states that are low on an ease-of-voting scale,  and report on its initial evaluation in a focus group. Our approach can be a win-win for voters, election agencies trying to fulfill their mandate and democracy at large.

\end{abstract}

%%
%% Keywords. The author(s) should pick words that accurately describe
%% the work being presented. Separate the keywords with commas.
% \keywords{Neural networks, Object Detection, Explainable AI, Adversarial AI}
% \keywords{Chatbots, Safety, Transparency, NLP, Voter Participation}

% ---------------------
\input{content/sec1_introduction}

\input{content/sec2_background}
\input{content/sec3_architecture}
\input{content/sec4_electionbot}

\input{content/sec5_discussion}
\input{content/sec6_conclusion}

% ---------------------

%%
%% The next two lines define the bibliography style to be used, and
%% the bibliography file.
%\bibliographystyle{format/aaai23}
\bibliography{references/biplav_trust.bib}
%\bibliography{references/biplav_trust.bib, references/kausik_ref.bib, references/election.bib, references/Shuge_rf.bib}

% ---------------------
%\appendix
%\input{content/appendix_rough}

% ---------------------
\end{document}

%% file: content/sec1_introduction.tex
\section{Introduction}

The success of any democracy depends on the ability of participants to vote in regular elections and the ability of the government to implement the subsequent orderly transfer of power. Traditionally, the act of voting has been conceptualized as the preeminent indicator of political participation \cite{voting-american,election-2016-voting}. Researchers have tried to estimate the effort needed to vote in different states in the United States using COVI, a cost of voting index \cite{election-cost,COVI}. Based on this index, Oregon ranks as the top state, where it is the easiest to vote in while New Hampshire ranks the last (50th). Within this spectrum, Mississippi and South Carolina, the two states we focus on, are ranked {{2nd}} and {{8th}} most complex states for citizens to vote in, respectively \cite{COVI}.  %(2nd and 8th)

When Artificial intelligence (AI) is referenced in connection with elections, it often draws negative reactions due to the fear of bots, misinformation, and hacking. This is particularly disappointing for chatbots, or bots, for short, since they are multi-modal collaborative assistants which have been studied since the early days of AI to help people complete useful tasks. For elections, people could overcome voting complexity by accessing information such as voting dates, jurisdiction, locations, and issues (propositions) on the ballot; and be informed on the voting process, equipment, and facilities at voting sites, conveniently in their own language or words through their smartphones, computers, and home devices like Alexa. Hence, in this paper, we set out to explore how chatbots  may be used to promote voter participation in vulnerable segments of society like senior citizens and first-time voters \cite{senior-elections}. 

We propose that AI technology can help voters get personalized information in a transparent manner from official government agencies whose mandate is to facilitate credible and inclusive elections.  To do so, we focus on the frequently asked questions (FAQs) that government agencies anticipate will be helpful to the voters. 
% Furthermore, we create a generic framework of safe and useful chatbots which can be more quickly and safely created than the state-of-the-% art by providing grounded answers, logging their interaction % with users, supporting common conversations, and being ac-% cessible. 
Furthermore, we develop a generic framework of {\em safe} and {\em useful} chatbots, which can be used to create chatbots more quickly and safely than the state-of-the-art \cite{survey-chatbot-use,survey-Thorne2017ChatbotsFT,apollo-chatbots}, with the following desirable properties - the chatbots developed will (i) only  provide {\em grounded answers} that can be traced to an official question/ answer; otherwise, the system is not confident of an answer and employs {\em "do not answer strategies"}, (ii) log their interaction with users so that the interaction can be audited, (iii) support a library of common conversation utterances, and (iv) be accessible via multiple modalities like speech. The properties (iii) and (iv) are for promoting usability.  

We use the developed framework to create election FAQ chatbots for two states in the United States - South Carolina (SC) and Mississippi (MS). We conducted an initial evaluation of one of the chatbots, i.e., for SC, in a focus group. Our approach can be a win-win for voters, election agencies trying to fulfill their mandate and democracy at large.

In the remainder of the paper, we give background about chatbots and elections in South Carolina and Mississippi. Then we present our reusable framework and two election chatbots for SC and MS.  We conclude with a discussions on the implications of this work and pointers for the future.

%% file: content/sec2_background.tex
\section{Background}

In this section, we give preliminaries about chatbots and election processes in the regions for which we consider the technology to improve voter participation. Election laws and procedures can vary greatly by state. In general, this makes it hard for any study to generalize nationally. We consider two different US states - South Carolina and Mississippi - to understand the supply, demand, and gap in voting information and how chatbots may help address the gap. This paper fortuitously overlaps with the 2022 general election in November, when local, state and national offices
will be decided.

% ----------
\subsection{Chatbots}

It is common to have real-world AI systems today that work with multimodal data like text, audio, and image. Some representative examples are recommender systems (for food, movies),  collaborative assistants (for health, bank transactions and travel) and route finders (for traffic navigation, emergency evacuation). Here, sample AI services for (a) natural language processing (NLP) are entity detection, text translation, summarization, and sentiment detection;  (b) audio processing include speech recognition and speaker identification; (c) image tasks such as object detection, face identification, object counting; and (d) other common tasks involving reasoning, search and ranking. 

However, the enthusiasm for AI services is also being dampened by  growing concerns about their reliability and trustworthiness with issues like bias, privacy and brittleness regardless of the data they use - be it text \cite{prob-bias-text}, audio \cite{prob-bias-sound} or image \cite{prob-bias-image}.  In fact, this has been argued as one of the key factors hampering the adoption of  AI techniques during an emergency like COVID-19 \cite{apollo-chatbots} or in safety-critical domains such as in medicine for treatments like breast cancer \cite{breastcancer-health-ai-uk}. In this context, we explore how the trustworthiness properties of AI could be assessed and communicated, without access to its code or data (i.e., in the black box setting), to promote its broader adoption.
% ----------
\subsection{South Carolina and its Elections}

% \bs{@bharath - say about the state and elections process, what is it electing in 2022}
% \bm{received this information from Brett}

% Researchers have tried to estimate the effort needed to vote in different states in the United States using COVI, a {\em cost of voting index} \cite{election-cost}. According to it, Oregon ranks as the top state where it is easiest to vote in while Texas ranks at the last (50th).  South Carolina has ranked the 7th most complex state for citizens to vote in, making it as a good candidate for studying how voting information is shared among residents.

In South Carolina, nearly 20\%
of the population in the state is age 65 and older (Census.gov, 2021). Elections are held in South Carolina to fill a number of municipal, state, and federal positions. These positions include Governor, Lieutenant Governor, Secretary of State, Attorney General, Treasurer, Comptroller General, Superintendent of Education, and Commissioner of Agriculture, as well as State Senate, State House of Representatives, State Delegation to the U.S. Senate, and State Delegation to the U.S. House of Representatives (Ballotpedia, 2022). The state’s election commission oversees voter registration, candidate certification, and elections (South Carolina Elections Commission, 2022). 
% According to \cite{election-cost}, %Schraufnagel (2020), 

South Carolina's official website has 30 FAQs \cite{sc-faqs},
% \footnote{Last accessed: 10/25/2022, \url{https://scvotes.gov/voters/voter-faq/}}, 
with average length of the questions being 10.9 words. The average answer length for the questions present in FAQs is 80.9 words. A total of 10 different topics are covered in the FAQ question-answer pairs. The topics include voting information, candidature information, absentee voting, polling information, ballot information, campaign, polling results, recounting, complaints, and general information. 
See summary in Table~\ref{tab:data_statistics}.

% ----------
\subsection{Elections in Mississippi}

% According to \cite{election-cost}, 
% Mississippi has ranked the 4th most complex state for citizens to vote in, making it another good candidate 
% %to providing an additional rationale 
% for studying how voting information is shared among residents.

In Mississippi, nearly 17\% of the
population in the state is aged 65 to 74 (Census.gov, 2021) but 76\% vote. In Fall 2022, Mississippi’s four House congressional races were all on the ballot.  The governor and the state’s two U.S. senators were not on the ballet, making this an “off” year with few big names.  Voter turnout was only 31.1\%, lower than the 2018 midterms (42.7\%) and the 2020 general election (59.9\%). Mississippi lacks some voting options available in many other states, including online voter registration. 

Mississippi's official website has 12 question-answer pairs available both in the website \cite{ms-faqs}
%\footnote{Last accessed: 11/11/2022, \url{https://www.sos.ms.gov/elections-voting/faqs}} 
and as a PDF document\footnote{ \url{https://tinyurl.com/4h22xb5e}}. The average length of the questions present in question-answer pairs is 7.75 words, with average length for the answers being 119.5 words. The number of topics covered in the FAQs is 5, belonging to absentee voting, voter registration, voting, information update, and candidature information. See summary in Table~\ref{tab:data_statistics}.

%% file: content/sec3_architecture.tex
\section{Safe and Usable Chatbot Architecture}

\begin{figure}[hbtp]
	\centering
	\includegraphics[width=0.47\textwidth,keepaspectratio]{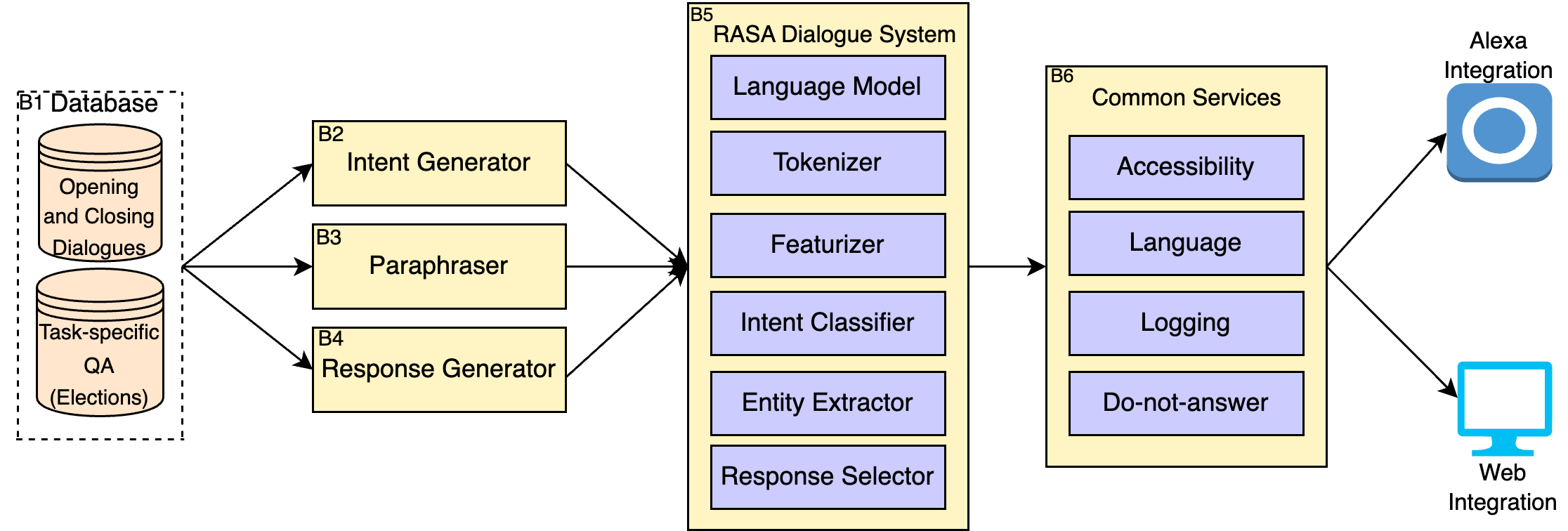}
	\caption{System Architecture}
	\label{fig:sys-arch}
\end{figure}

We now present a general architecture for building chatbots with the goals of promoting safety  and usability (Figure ~\ref{fig:sys-arch}). For safety, the chatbot maintains a list of verified
questions and their answers, and only responds to use questions whose intent it has recognized with high confidence and when it has been able to determine {\em grounded answers} that can be traced to the official
question/ answer list. Second is
logging the system's interaction with the users so that the interaction
can later be audited.
For usability, the framework maintains a library of  common conversation utterances for opening and closing the chat which any chatbot implementation can reuse, and enables created chatbots to be accessible via additional modalities like speech via Alexa.

% Features that make it safe are:
% \begin{itemize}
%     \item Logging of conversations
%     \item Do not answer strategies
%     \item Paraphrasing for better question matching
%     \item Alexa Integration
% \end{itemize}

Figure \ref{fig:sys-arch} shows our proposed system architecture. Key components of our system are:
\begin{itemize}
\item {\bf Database (B1)}: The database is the source from which we extract the training data to train the chatbot. We ensure that the source is reliable and trustworthy. Task-specific QA refers to the data source pertaining to the chosen domain (which is elections, in this case). The opening and closing dialogues are usually generic (like greeting and saying bye).
\item {\bf Intent Generator (B2)}: Intent Generation is crucial in building an efficient and functional chatbot that will determine the bot's success in fulfilling the user's needs. Intent Generator helps in tagging existing questions to an intent, which can later be utilized to map any new incoming user utterance to an available intent to provide desired answers. Many off-the-shelf intent generator modules are available such as RASA's DIET \cite{bunk2020diet}. However, intent generator modules can be generated as per the domain under consideration using various approaches ranging from word tagging to deep learning.
\item {\bf Paraphraser (B3)}: In order to train a chatbot, it is crucial to present it with similar questions that match an answer, rather than having only a single question associated with an answer. However, in any FAQ wesbite, we only obtain a single question tagged with an answer. Thus, in order to create an efficient chatbot, we made use of a paraphrasing tool to generate similar questions as to the one extracted from the election FAQ website. 
\item {\bf Response Generator (B4)}: Responses are messages that the chatbot sends to the user. A response is usually text, but can also include multi-modal content like images and audio. The safe chatbot architecture reuses the response generation module available in the RASA Dialogue System. This module helps provide various ways to respond to a user's utterance using Custom Actions, Response Variations, Conditional Response Variations, Images, and Rich Response Buttons among many others.
\item {\bf RASA Dialogue System (B5)}: We use the RASA chatbot framework \cite{rasa} to build the chatbot. The dialogue system has an NLU pipeline with different components for understanding human conversation and responding appropriately. Language models like the Spacy language model can be used if one wants to use pre-trained word vectors. Tokenizer converts sentences to tokens. Featurizer creates a vector representation of the user message and response. The intent classifier classifies the intent of the user message. The entity extractor extracts entities that are specified in the training data. The response selector chooses the appropriate response based on the identified intent and entities. 
\item {\bf Common Services (B6)}: The common services are optional and the user has the flexibility of choosing the services they need. Some of the accessibility options are font settings and Text-to-Speech. The users will be able to converse with the chatbot in a language that is comfortable for them using the language settings. This can be implemented by making use of translators. The conversations can be logged for storage and retrieval using the logging option. This also helps the developers to improve the chatbot conversation by reviewing the stored conversations. We do not want the bot to respond to certain questions (for example, questions like 'Do you think the current president is doing a good job?'). These questions are mostly subjective. The 'Do-not-answer' option can be used in this case. 

\item {\bf System Integration}: Web integration and Alexa integration provide an engaging user interface to converse with the chatbot. We integrated the RASA chatbot with the Alexa device as a skill using the Alexa developer console (the skill is still in the beta phase). Web integration was done using the RASA webchat package. The system built for South Carolina can be accessed here \url{http://casy.cse.sc.edu/Election-Chatbot-SC-main/} and the system built for Mississippi is available at \url{http://casy.cse.sc.edu/ElectionBot-MS-main/Chatbot/}. 
\end{itemize}

Our architecture has been built, generalizing our experience building chatbots across different domains where the usability of the system is essential (for example, education \cite{allure}, networking \cite{nl2sql}, etc.) with little or no modifications. This shows the generalizability of our system. At the same time, common services represent safety-specific services we have implemented to ensure that the chatbot's behavior can be audited, controlled, and made more accessible to users. Logging makes the conversations more accessible and helps developers review the conversations to ensure that the interaction flows safely. The 'Do-not-answer' option disables the chatbot from answering inappropriate questions. The blocks B2, B3 and B4 also work in a controlled manner and we only use reliable sources (like Government websites) for B1. Based on our past experience, all these features benefit many other applications where user safety is essential. 
% \bs{Kausik - we want to say that our system has specific features based on our experience, and it will benefit many other applications where user safety is important.} \kl{Done. Please review.}

% \bs{Chatbot - reuse, testing, usability}

\subsection{Chatbot Testing}
For a user to trust a chatbot, it should be competent and reliable. The performance of chatbots can be evaluated using many methods such as user studies with focus groups or by administering surveys, A/B testing, and Randomized Controlled Trials (RCTs). Chatbot testing allows us to improve the chatbot design and functionality.

\subsubsection{Focus Groups:} are a type of in-depth scientific study where a small group of participants with the desired backgrounds is invited to engage with a technology and structured as well as unstructured questions could be asked. Typically, the sessions are recorded and  both qualitative and quantitative analyses could be conducted after the session.

\subsubsection{Randomized Controlled Trials (RCTs):} are a form of a scientific experiment that is used to test the effectiveness of a new treatment. In this case, it would be a new chatbot that is being tested. It is a study in which the participants are randomly assigned to either a treatment or control group. Participants in the control group receive an alternative treatment (placebo). The two groups are compared to test the effectiveness of the new chatbot (treatment). Though RCTs provide a way to test the chatbots with minimal or no bias and confounding effects, they require time, money, and many participants. They have been widely used for evaluating medical devices and vaccines technologies in critical domains like medicine, and sometimes for chatbots\cite{woebot}, but not commonly for technology during elections, with some exceptions \cite{voter-chatbot-rct}.  

\subsubsection{RCTs in safe chatbot framework:} Trustworthiness in answers and the reliability of the users' experiences are parts of the essential components of a safe chatbot. We propose to use RCTs to test the attributes; in this case, we test the effectiveness of answers from the chatbot. We build two versions of chatbots. One is the chatbot being tested, called the experimental chatbot; another one is the placebo chatbot, called the control chatbot. The experimental chatbot is the one built using the mentioned architectures with prepared Q-As in the FAQ list. The control chatbot will send the web pages where the answers reside instead of the specific answers. Users are randomly assigned to these two chatbots and are asked to give feedback on every question being answered. The RCT experiment can be conducted before the chatbot is put into practice. We can analyze whether there are statistical differences in each question's feedback between the two chatbots. The chatbot is effective only when its feedback scores are not lower than the control chatbot. Otherwise, there is no justified need for the chatbot. Other attributes (Activation Rate, Fall Back Rate, Retention Rate, Self-Service Rate, Confusion Triggers, etc.) can also be tested in the same ways by using respective feedback measurements \cite{xu2020recipes,safe-xu-etal-2021-bot}. 

%% file: content/sec4_electionbot.tex
\begin{figure*}[h!]
    \centering
    \includegraphics[width=0.7\linewidth,keepaspectratio]{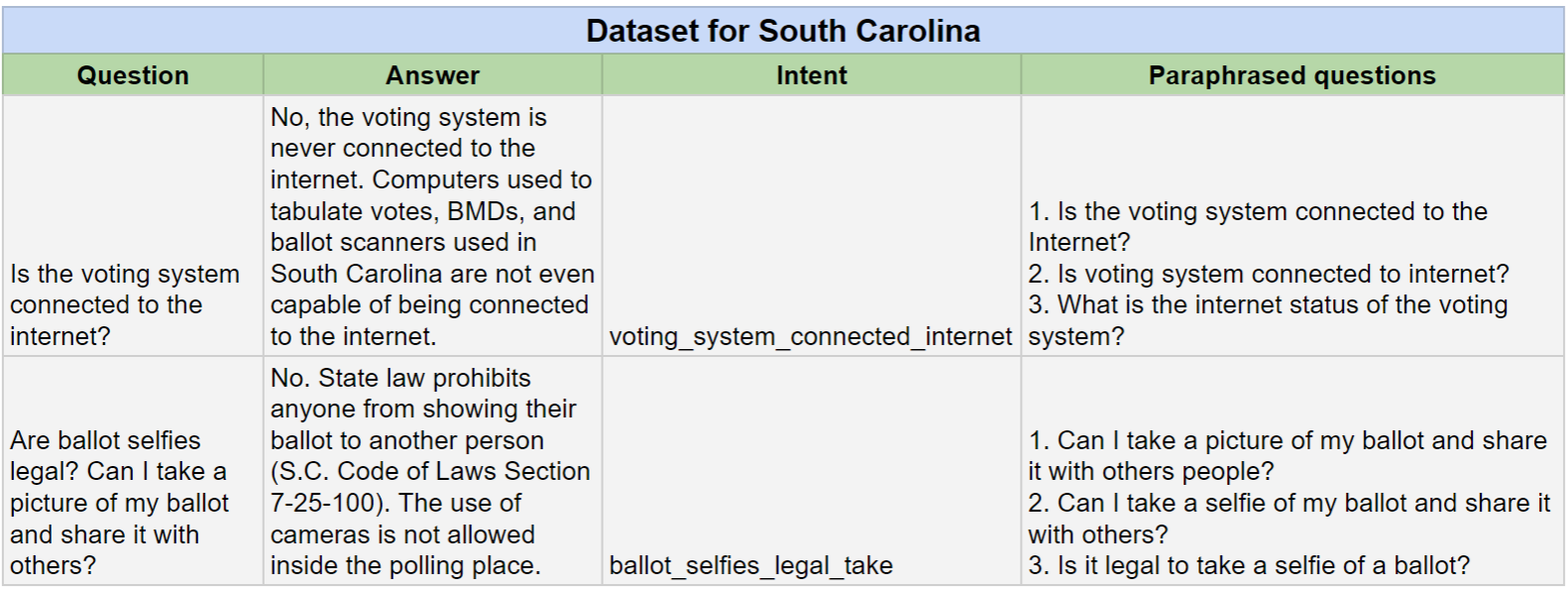}
    \caption{A snapshot of internal representation for South Carolina data.}
    \label{fig:dataset_sc}
\end{figure*}

\begin{figure*}[h!]
    \centering
    \includegraphics[width=0.7\linewidth,keepaspectratio]{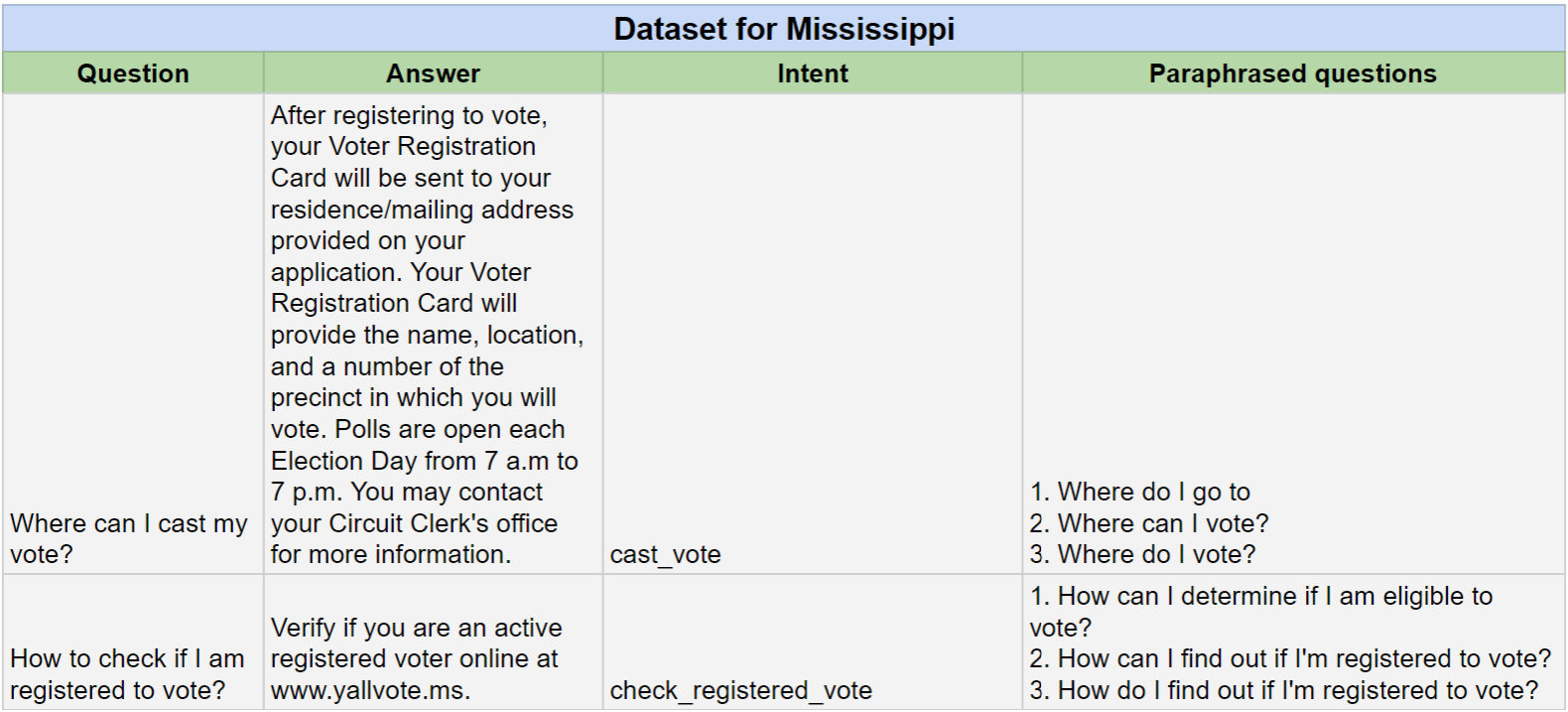}
    \caption{A snapshot of internal representation for Mississippi data.}
    \label{fig:dataset_ms}
\end{figure*}

% -----
\section{Election Chatbot Implementation}

In this section, we describe the implementation details for election chatbots designed to answer frequently answered questions for the states of Mississippi and South Carolina. We describe in detail the procurement of datasets required for these chatbots, their design, and assessment of the developed chatbots.

% \bs{Assuming we keep a single section, we need a table to compare and contrast SC with MS. Its rows can be the dimensions like data, Q/A features, gaps, assessment. Also, we need to have SC and MS specific statements separately under clear headings. }

\begin{table}[h]
    \centering
    \begin{tabular}{|l|c|c|}
    \hline
     & \textbf{South Carolina} & \textbf{Mississippi} \\ \hline
    Number of Q/A pairs & 30 & 12 \\ \hline
    Avg question length & 10.9 & 7.75 \\ \hline
    Avg answer length & 80.9 & 119.5 \\ \hline
    Number of topics & 10  &  5 \\ \hline
    Last updated & Oct 25,2022  & Nov 11,2022 \\ \hline
    \end{tabular}
    \caption{Statistics about FAQ  Data. Question and answer lengths are measured in words.}
    \label{tab:data_statistics}
\end{table}

\subsection{Data for Q/A}

For building the datasets to train the election chatbots, we have made use of the frequently answered questions (FAQs) available on the government websites pertaining to elections from respective states (South Carolina \cite{sc-faqs}, Mississippi \cite{ms-faqs}). The intuition behind using an official government website to procure question-answer (Q-A) pairs is to inherently keep the user interaction controlled and secure.

We have used a web-scraping tool in order to automatically extract the FAQs and save them in the form of comma-seperated values (.csv) files. Table~\ref{tab:data_statistics} shows a comparative summary of the FAQs while Figure \ref{fig:dataset_sc} and Figure \ref{fig:dataset_ms} show the snapshot of Q-A pair datasets built for South Carolina and Mississippi. For South Carolina, we have obtained 30 Q-A pairs, and 12 Q-A pairs for Mississippi. Notably, FAQ questions in MS are fewer, cover  fewer topics and  answers are longer than SC. 

% {\color{red} \noindent {\bf SC considerations:}

% \noindent {\bf MS considerations:}}

\subsection{Chatbot Design and Implementation}
With the datasets obtained for the two state under study, we now utilize it to design the election chatbots for South Carolina and Mississippi. In order to do so, we reuse the safe chatbot architecture described in the previous section.

\noindent {\bf SC considerations:}
Reused, without change. \\
\noindent {\bf MS considerations:}
Reused, without change.

\noindent \textbf{Intent Generator:} Questions in the dataset for both South Carolina and Mississippi are small, with an average of ~7 words in a sentence. On further exploration, we identified that most of these words are stop-words, i.e., commonly used words in any language. For example, in English, "the", "is", and "why" qualify as stop-words. In order to obtain important information, it is important to remove stop-words, and when we carried out the similar process on the election FAQ datasets, we identified that there are only a few words left out for every Q-A pair. Thus, in order to obtain the intent for each question, we have used n-gram approach to tag each question with an intent consisting of the important words present in the initially extracted questions. 

\noindent \textbf{Paraphraser:} We make use of a paraphrasing tool available on Huggingface\footnote{https://huggingface.co/eugenesiow/bart-paraphrase}, which is a repository consisting of trained language models for inference. We generate 3 similar sentences to the questions present in both FAQ datasets for SC and MS.  

\noindent \textbf{Integration with RASA:}
With the trained dataset consisting of paraphrasing questions for each state under consideration, we train the RASA bot to generate election chatbots corresponding to SC and MS. Having a RASA chatbot helps ease the deployment process as well as there is a mechanism offered by RASA for multiple deployment opportunities.

\noindent \textbf{Deployment}
The developed election chatbots for SC and MS are deployed on server using Django\footnote{https://www.djangoproject.com/} and Amazon Alexa. The websites for election chatbots belonging to SC and MS facilitate user interaction through the medium of typed conversations. On the other hand, Alexa offers the medium of voice-based conversations with the election chatbots designed for SC and MS. Deploying the chatbots for public usage assists in aggressive testing of the chatbots.

% -----
\subsection{User Assessment}

We planned an initial  evaluation of the chatbots in focus group settings and then move to larger approaches like surveys. The reason for this is that we anticipate lots of  questions about why certain answers were given (or not) and could explain the context of the chatbot's behavior. Further, given the socio-technical nature of the problem and multiple agencies involved in elections, we wanted to ensure that the results are not misunderstood.

\noindent {\bf SC considerations:}

We conducted an initial study with 
two focus groups\footnote{IRB Exemption and Compensation: This research study has been certified as exempt from the IRB per 45 CFR 46.104(d)(3) and 45 CFR 46.111(a)(7) by University of South Carolina IRB\# Pro00124482.  Participants were not compensated for participating in the focus group.}. The first focus group was on Oct 28, 2022 and had 
two users who were senior citizens and long-time voters. The users noted that in the context of elections, the most common information they seek is about {\em time}, {\em place} and {\em procedures} for participating in elections like absentee ballots and working of election equipments.  The second focus group was on Dec 7, 2022 at a senior citizen center and had four participants. The group identified additional categories about which they seek information  as {\em candidate backgrounds} and {\em propositions on referendum}. 
% ...
% \bs{Bharath - work with Brett on a summary}. 
% The users noted that in the context of elections, the most common information they seek is about {\em time}, {\em place} and {\em procedures} for participating in elections like absentee ballots.  Among these, only information about election procedures were available in the SC FAQs \cite{sc-faqs}.

\noindent {\bf MS considerations:}
We conducted an initial study with 
one focus group\footnote{IRB Exemption and Compensation: This research study has been certified as exempt from the IRB per 45 CFR 46.104(b)(\#2)  by University of Mississippi IRB\# 23x-100. Participants were not compensated for participating in the focus group.}.
In this initial study, we tested the chatbot on November 17, 2022 with one user, a senior citizen, and a long-time voter. The user noted that finding information regarding voter registration, requirements for voting, and the nearest vote center is of utmost importance for senior citizens. The user also emphasized absentee voting, as many senior citizens might be unable to travel to the voting center due to physical considerations. After interaction with the chatbot, the user expressed that the preferred medium for information consumption regarding voting is in-person and through government officials. It was also noted that only a few senior citizens have access to technology or the expertise to interact with it.

\noindent {\bf Summary:} From the focus groups in two different states with senior citizens, some of whom have eye, hearing and other cognitive disabilities, we found they wanted  decision support tools provided they give useful information. They identified 5 categories from which they seek information but the two states only had information from 1 category ({\em procedures}). 
Hence, the current situation is unsatisfactory and the need for technology-based voter enablement is real from the collected evidence.
One  could consider non-official information sources but then the technology developer cannot assure information quality in general. One way out may be to give usable information from all categories but provide disclaimer and contextualize the information with its risks.

%% file: content/sec5_discussion.tex
\section{Discussion}  

In previous sections, we discussed the design of a reusable decision support     framework in natural language, i.e., a chatbot or collaborative assistant, for providing answers to frequently asked questions safely. We also implemented two chatbots using it to provide FAQ information for voting in in South Carolina and Mississippi, which can be easily scaled for other states. The safety  highlights of system are its ability to only answer questions that are grounded and support post-facto auditing, and usability features are para-phrasing of questions, a rich library of opening and closing conversations and integration for text and speech based interaction

Dialogue safety is important to the conversational system, especially when the dialogue systems are trained on open-domain datasets or based on large pre-trained language models. The proposed voting chatbots are closed-domain chatbots and hence, will not suffer the same risks as dialogue models trained on the open-domain corpora. Other significant trust issues with chatbots relate to usage of abusive language, leaking information and usage of complex terms \cite{chatbot-ethical}. There are many emerging techniques to handle them  and in particular, rating  chatbot for trustworthiness based on automated independent testing, so that users are aware of the chatbot behavior \cite{chat-rating}.

% To illustrate, the closed-domain voting chatbots only use the "Do Not Answer" intents and refuse to answer those pre-setting sensitive questions such as "who will win the election?", the chatbots will fail to avoid toxic questions that are not listed in the "Do Not Answer" intents. The "Do Not Answer" is safe but far from engaging. There is often a trade-off between providing safe and grounded answers that can be traced back to official FAQs (provenance, transparency) and user engagement. 
% More comprehensive safety strategies for close-domain chatbots can be added to the system. 

Domain-specific "Do Not Answer" prototypes are yet to be explored at a large scale. For domain-specific agents, like voting systems, healthcare, and legal bots, there is no open corpus for these dialogue systems. Each domain or even each bot has its specific unsafe dialogues. This paper uses a simple but flexible approach, "Do not Answer" intents, to allow adding as many as known unsafe dialogues. Yet it can not deal with unknown unsafe questions. A more exhaustive domain-specific safe training corpus is yet to be explored. 

% Applying the structure to other information chatbots is further to be studied. This paper explores putting the website information into a chatbot, which makes it more straightforward to get the information. Many websites are also there for publishing information, such as the university website for student service. Any other websites that receive a high quantum of inquiries on FAQs have the potential to use the effortless-building safe dialogue system. Yet more application scenarios are to be explored.

% What can be further improved is to build a  chatbot at the national level, i.e., "United States" voting information chatbot, that can host voting information for the all states in a single chatbot. The current architecture only holds one state in a single chatbot. Will the united chatbot respond to the questions from South Carolina with the Mississippi answers? It will be further explored and one chatbot for all voting information seems more practical for the public.

%% file: content/sec6_conclusion.tex
\section{Conclusion}  

In this paper, we showed that AI can play a positive role in driving voter participation.
We considered the challenges vulnerable sections of the electorate face during voting, and positioned chatbots to address
those challenges in a trustworthy manner. 

There are many avenues for the future. One could extend the range of information that the chatbot (collaborative assistant) can offer while transparently showing where it got the results from. Second, one could build chatbots for  more states and evaluate extensively using RCTs. Third, one could extend the multi-modal capabilities of the chatbot to be more engaging.

%% file: main.bbl
\begin{thebibliography}{24}
\providecommand{\natexlab}[1]{#1}

\bibitem[{Almansor and Hussain(2020)}]{survey-chatbot-use}
Almansor, E.; and Hussain, F. 2020.
\newblock \emph{Survey on Intelligent Chatbots: State-of-the-Art and Future
  Research Directions}, 534--543.
\newblock ISBN 978-3-030-22353-3.

\bibitem[{Antun et~al.(2020)Antun, Renna, Poon, Adcock, and
  Hansen}]{prob-bias-image}
Antun, V.; Renna, F.; Poon, C.; Adcock, B.; and Hansen, A.~C. 2020.
\newblock On instabilities of deep learning in image reconstruction and the
  potential costs of AI.
\newblock \emph{Proceedings of the National Academy of Sciences}, 117(48):
  30088--30095.

\bibitem[{Blodgett et~al.(2020)Blodgett, Barocas, au2, and
  Wallach}]{prob-bias-text}
Blodgett, S.~L.; Barocas, S.; au2, H. D.~I.; and Wallach, H. 2020.
\newblock Language (Technology) is Power: A Critical Survey of "Bias" in NLP.
\newblock In \emph{On Arxiv at: 2https://arxiv.org/abs/2005.14050}.

\bibitem[{Bocklisch et~al.(2017)Bocklisch, Faulkner, Pawlowski, and
  Nichol}]{rasa}
Bocklisch, T.; Faulkner, J.; Pawlowski, N.; and Nichol, A. 2017.
\newblock Rasa: Open Source Language Understanding and Dialogue Management.

\bibitem[{Bunk et~al.(2020)Bunk, Varshneya, Vlasov, and Nichol}]{bunk2020diet}
Bunk, T.; Varshneya, D.; Vlasov, V.; and Nichol, A. 2020.
\newblock Diet: Lightweight language understanding for dialogue systems.
\newblock \emph{arXiv preprint arXiv:2004.09936}.

\bibitem[{COVI(June 2022)}]{COVI}
COVI. June 2022.
\newblock Cost of Voting Index, Values and Rankings.
\newblock \url{https://costofvotingindex.com}.
\newblock Accessed: 2022-11-11.

\bibitem[{Henderson et~al.(2018)Henderson, Sinha, Angelard{-}Gontier, Ke,
  Fried, Lowe, and Pineau}]{chatbot-ethical}
Henderson, P.; Sinha, K.; Angelard{-}Gontier, N.; Ke, N.~R.; Fried, G.; Lowe,
  R.; and Pineau, J. 2018.
\newblock Ethical Challenges in Data-Driven Dialogue Systems.
\newblock In Furman, J.; Marchant, G.~E.; Price, H.; and Rossi, F., eds.,
  \emph{Proceedings of the 2018 {AAAI/ACM} Conference on AI, Ethics, and
  Society, {AIES} 2018, New Orleans, LA, USA, February 02-03, 2018}, 123--129.
  {ACM}.

\bibitem[{Koenecke et~al.(2020)Koenecke, Nam, Lake, Nudell, Quartey, Mengesha,
  Toups, Rickford, Jurafsky, and Goel}]{prob-bias-sound}
Koenecke, A.; Nam, A.; Lake, E.; Nudell, J.; Quartey, M.; Mengesha, Z.; Toups,
  C.; Rickford, J.~R.; Jurafsky, D.; and Goel, S. 2020.
\newblock Racial disparities in automated speech recognition.
\newblock \emph{Proceedings of the National Academy of Sciences}, 117(14):
  7684--7689.

\bibitem[{Lakkaraju et~al.(2022{\natexlab{a}})Lakkaraju, Hassan, Khandelwal,
  Singh, Bradley, Shah, Agostinelli, Srivastava, and Wu}]{allure}
Lakkaraju, K.; Hassan, T.; Khandelwal, V.; Singh, P.; Bradley, C.; Shah, R.;
  Agostinelli, F.; Srivastava, B.; and Wu, D. 2022{\natexlab{a}}.
\newblock ALLURE: A Multi-Modal Guided Environment for Helping Children Learn
  to Solve a Rubik’s Cube with Automatic Solving and Interactive
  Explanations.
\newblock \emph{Proceedings of the AAAI Conference on Artificial Intelligence},
  36(11): 13185--13187.

\bibitem[{Lakkaraju et~al.(2022{\natexlab{b}})Lakkaraju, Palaiya, Paladi,
  Appajigowda, Srivastava, and Johri}]{nl2sql}
Lakkaraju, K.; Palaiya, V.; Paladi, S.~T.; Appajigowda, C.; Srivastava, B.; and
  Johri, L. 2022{\natexlab{b}}.
\newblock Data-Based Insights for the Masses: Scaling Natural Language Querying
  to Middleware Data.
\newblock In \emph{Database Systems for Advanced Applications: 27th
  International Conference, DASFAA 2022, Virtual Event, April 11–14, 2022,
  Proceedings, Part III}, 527–531. Berlin, Heidelberg: Springer-Verlag.
\newblock ISBN 978-3-031-00128-4.

\bibitem[{Mann(2021)}]{voter-chatbot-rct}
Mann, C.~B. 2021.
\newblock {Can Conversing with a Computer Increase Turnout? Mobilization Using
  Chatbot Communication}.
\newblock \emph{Journal of Experimental Political Science}, 8(1): 51--62.

\bibitem[{Mississippi(2022)}]{ms-faqs}
Mississippi. 2022.
\newblock Election Frequently Asked Questions.
\newblock \emph{https://www.sos.ms.gov/elections-voting/faqs, last accessed Nov
  11}.

\bibitem[{Narayanan et~al.(2021)Narayanan, Robertson, Hickerson, Srivastava,
  and Smith}]{senior-elections}
Narayanan, V.; Robertson, B.~W.; Hickerson, A.; Srivastava, B.; and Smith,
  B.~W. 2021.
\newblock Securing social media for seniors from information attacks: Modeling,
  detecting, intervening, and communicating risks.
\newblock In \emph{The Third IEEE International Conference on Cognitive Machine
  Intelligence (IEEE CogMI)}.

\bibitem[{PewResearchCenter(2018)}]{election-2016-voting}
PewResearchCenter. 2018.
\newblock An examination of the 2016 electorate, based on validated voters.

\bibitem[{Prochaska et~al.(2021)Prochaska, Vogel, Chieng, Kendra, Baiocchi,
  Pajarito, and Robinson}]{woebot}
Prochaska, J.~J.; Vogel, E.~A.; Chieng, A.; Kendra, M.; Baiocchi, M.; Pajarito,
  S.; and Robinson, A. 2021.
\newblock A Therapeutic Relational Agent for Reducing Problematic Substance Use
  (Woebot): Development and Usability Study.
\newblock \emph{J Med Internet Res}, 23(3): e24850.

\bibitem[{Schraufnagel, Pomante~II, and Li(2020)}]{election-cost}
Schraufnagel, S.; Pomante~II, M.~J.; and Li, Q. 2020.
\newblock Cost of Voting in the American States: 2020.
\newblock \emph{Election Law Journal: Rules, Politics, and Policy}, 19(4):
  503--509.

\bibitem[{Seedat and Taylor-Phillip(2022)}]{breastcancer-health-ai-uk}
Seedat, F.; and Taylor-Phillip, S. 2022.
\newblock UK guidance on evaluating AI for use in breast screening.
\newblock In
  \emph{https://nationalscreening.blog.gov.uk/2022/08/01/guidance-on-evaluating-ai-for-use-in-breast-screening/}.

\bibitem[{South-Carolina(2022)}]{sc-faqs}
South-Carolina. 2022.
\newblock Election Frequently Asked Questions.
\newblock \emph{https://scvotes.gov/voters/voter-faq, last accessed Oct 25}.

\bibitem[{Srivastava(2021)}]{apollo-chatbots}
Srivastava, B. 2021.
\newblock Did chatbots miss their “Apollo Moment”? Potential, gaps, and
  lessons from using collaboration assistants during COVID-19.
\newblock In \emph{Patterns, Volume 2, Issue 8, 100308}.

\bibitem[{Srivastava et~al.(2020)Srivastava, Rossi, Usmani, and
  Bernagozzi}]{chat-rating}
Srivastava, B.; Rossi, F.; Usmani, S.; and Bernagozzi, M. 2020.
\newblock Personalized Chatbot Trustworthiness Ratings.
\newblock In \emph{IEEE Transactions on Technology and Society.}

\bibitem[{Thorne(2017)}]{survey-Thorne2017ChatbotsFT}
Thorne, C. 2017.
\newblock Chatbots for troubleshooting: A survey.
\newblock \emph{Lang. Linguistics Compass}, 11.

\bibitem[{Verba and Nie(1972)}]{voting-american}
Verba, S.; and Nie, N.~H. 1972.
\newblock Participation in America: Political democracy and social equality.
\newblock \emph{New York: Harper Row.}

\bibitem[{Xu et~al.(2020)Xu, Ju, Li, Boureau, Weston, and
  Dinan}]{xu2020recipes}
Xu, J.; Ju, D.; Li, M.; Boureau, Y.-L.; Weston, J.; and Dinan, E. 2020.
\newblock Recipes for safety in open-domain chatbots.
\newblock \emph{arXiv preprint arXiv:2010.07079}.

\bibitem[{Xu et~al.(2021)Xu, Ju, Li, Boureau, Weston, and
  Dinan}]{safe-xu-etal-2021-bot}
Xu, J.; Ju, D.; Li, M.; Boureau, Y.-L.; Weston, J.; and Dinan, E. 2021.
\newblock Bot-Adversarial Dialogue for Safe Conversational Agents.
\newblock In \emph{Proceedings of the 2021 Conference of the North American
  Chapter of the Association for Computational Linguistics: Human Language
  Technologies}, 2950--2968. Online: Association for Computational Linguistics.

\end{thebibliography}
